\documentclass[showpacs,prl,onecolumn,aps,superscriptaddress,preprintnumbers,nofootinbib]{revtex4}
\usepackage[T1]{fontenc}
\usepackage[latin9]{inputenc}
\usepackage{amsmath,amssymb}
\usepackage{epsfig}
\usepackage{bm} 
\usepackage{graphicx}
\usepackage{mathrsfs}
\usepackage{amsmath}
\usepackage{amsfonts}
\usepackage{epstopdf}
\usepackage{color}
\usepackage{pifont}
\usepackage{relsize}
\usepackage{mathtools}
\def\slashchar#1{\setbox0=\hbox{$#1$}     		
   \dimen0=\wd0                                 	
   \setbox1=\hbox{/} \dimen1=\wd1               	
   \ifdim\dimen0>\dimen1                        	
      \rlap{\hbox to \dimen0{\hfil/\hfil}}      	
      #1                                        	
   \else                                        	
      \rlap{\hbox to \dimen1{\hfil$#1$\hfil}}   	
      /                                         	
   \fi}

\renewcommand{\vec}{\boldsymbol}
\newcommand{\beq}{\begin{equation}}
\newcommand{\eeq}{\end{equation}}
\newcommand{\bea}{\begin{eqnarray}}
\newcommand{\eea}{\end{eqnarray}}
\newcommand{\ba}{\begin{array}}
\newcommand{\ea}{\end{array}}

\def\eq#1{{Eq.~(\ref{#1})}}
\def\fig#1{{Fig.~\ref{#1}}}
\newcommand{\bas}{\bar{\alpha}_S}
\newcommand{\as}{\alpha_S}
\newcommand{\nn}{\nonumber}

\newcommand{\Lb}{\left(}
\newcommand{\Rb}{\right)}
\newcommand{\h}{\frac{1}{2}}

\newcommand{\zz}{\tilde z}
\newcommand{\intl}{\int\limits}
\begin{document}

\title{Multiplicity distribution and entropy of produced gluons in  deep inelastic scattering at high energies}

\author{Eugene Levin}
\email{leving@tauex.tau.ac.il, eugeny.levin@usm.cl}
\affiliation{Department of Particle Physics, Tel Aviv University, Tel Aviv 69978, Israel}
\affiliation{Departemento de F\'isica, Universidad T\'ecnica Federico Santa Mar\'ia, and Centro Cient\'ifico-\\
Tecnol\'ogico de Valpara\'iso, Avda. Espana 1680, Casilla 110-V, Valpara\'iso, Chile}

\date{\today}


\begin{abstract}
In this paper we found the multiplicity distribution of the produced gluons in deep inelastic scattering at large $z=\ln\Lb Q^2_s/Q^2\Rb\,\,\gg\,\,1$ where $ Q_s $ is the saturation  momentum and $Q^2$ is the photon  virtuality.
It turns out that this distribution  at large $n >  \bar{n}$ almost reproduces   the KNO scaling behaviour with the average number of gluons $ \bar{n} \propto \exp\Lb z^2/2 \kappa\Rb$, where $\kappa = 4.88 $ in the leading order of perturbative QCD.   The KNO function $\Psi\Lb \frac{n}{\bar{n}}\Rb = \exp\Lb -\,n/\bar{n}\Rb$.
For $n < \bar{n}$ we found that $\sigma_n \propto \Big( z - \sqrt{2 \,\kappa\,\ln (n-1)}\Big)/(n-1)$.  Such small $n$ determine the value of entropy of produced gluons $S_E = 0.3\, z^2/(2\,\kappa)$ at large $z$.  The factor $0.3$ stems from the non-perturbative corrections that provide the correct behaviour of the saturation momentum at large $b$.

 \end{abstract}
\maketitle

\vspace{-0.5cm}
\tableofcontents

\section{Introduction}
 Deep inelastic scattering (DIS) processes  play a unique role in developing the theoretical understanding of the high energy interaction in QCD. Indeed, 
 the main ideas of Colour Glass Condensate(CGC)/saturation approach (see  Ref.\cite{KOLEB} for a review) : the saturation of the dipole density and the new dimensional scale ($Q_s$), which increases with energy, have become the common language for discussing the high energy scattering in QCD. All these ideas are rooted in the theoretical description of DIS\cite{BFKL,LIP,LIREV,GLR,MUQI,MUDI,MV}.
 
We are going to discuss the multiplicity distribution of the produced gluons in DIS.  The multiplicity distributions and especially the entropy of produced gluons have become a hot subject during the past several years   
\cite{KUT,PES,KOLU1,PESE,KHLE,BAKH,BFV,HHXY,KOV1,GOLE1,GOLE2,KOV2,NEWA,LIZA,FPV,TKU,KOV3,KOV4,DVA1}  and we hope that this paper will look at these problems at different angle. However, before discussing the multiplicity distribution it looks reasonable to summarize  our theoretical achievements   in DIS. 

First and most importantly is   that  the scattering amplitude of the colourless dipole with the size $x_{01}$ which determines the DIS cross section, satisfies the Balitsky-Kovchegov (BK) non-linear equation\cite{BK}:
 \beq \label{BK}
 \frac{\partial N_{01}}{\partial Y}\,=\,\bas\int \frac{d^2\,x_{02}}{2 \pi} \frac{ x^2_{01}}{x^2_{02}\,x^2_{12}}\Big\{ N_{02} + N_{12} - N_{02}N_{12} - N_{01}\Big\}
 \eeq
 where $N_{ik}=N\Lb Y, \vec{x}_{ik},\vec{b}\Rb$ is the scattering amplitude of the dipoles with size $x_{ik}$ and with rapidity $Y$ at the impact parameter $\vec{b}$.
 
 It has been shown that \eq{BK} leads to a new dimensional scale: saturation momentum\cite{GLR}  which has the following $Y$ dependence\cite{GLR,MUT,MUPE}:
 \beq \label{QS}
 Q^2_s\Lb Y, b\Rb\,\,=\,\,Q^2_s\Lb Y=Y_0, b\Rb \,e^{\bas\,\kappa \,Y\,-\,\,\frac{3}{2\,\gamma_{cr}} \ln Y }
 \eeq 
 where $Y_0$ is the initial value of rapidity and $\kappa$ and $\gamma_{cr}$   are determined by the following equations\footnote{$\chi\Lb \gamma\Rb$ is the BFKL kernel\cite{BFKL} in anomalous dimension ($\gamma$) representation.$\psi$ is the Euler psi -function (see Ref.\cite{RY} formula {\bf 8.36}). }:
 
 \beq \label{GACR}
\kappa \,\,\equiv\,\, \frac{\chi\Lb \gamma_{cr}\Rb}{1 - \gamma_{cr}}\,\,=\,\, - \frac{d \chi\Lb \gamma_{cr}\Rb}{d \gamma_{cr}}~~~\,\,\,\mbox{and}\,\,\,~~~\chi\Lb \gamma\Rb\,=\,\,2\,\psi\Lb 1 \Rb\,-\,\psi\Lb \gamma\Rb\,-\,\psi\Lb 1 - \gamma\Rb
\eeq
  
 In Ref.\cite{MUT} (see also Re,\cite{MUPE}) it is shown that in the vicinity  of the saturation scale the scattering amplitude takes the following form: 
  \beq \label{VQS}
 N_{01}\Lb  z\Rb\,\,\,=\,\,\,\mbox{Const} \Lb x^2_{10}\,Q^2_s\Lb Y\Rb\Rb^{\bar \gamma}
 \eeq
 with $\bar \gamma = 1 - \gamma_{cr}$. In \eq{VQS} we introduce a new variable $z$, which is equal to:
  \beq \label{z}
 z\,\,=\,\,\ln\Lb x^2_{01}\,Q^2_s\Lb Y, b\Rb\Rb\,\,\, =\,\,\,\,\bas\,\kappa \,\Lb Y\,-\,Y_A\Rb\,\,+\,\,\xi
 \eeq  
 where $\xi = \ln x^2_{01}$. Note that  we neglected the term $\frac{3}{2\,\gamma_{cr}} \ln Y$ in \eq{QS}. 
 
 It turns out that inside the saturation region: $x^2_{01}\,Q^2_s\Lb Y\Rb \,>\,1$ the scattering amplitude shows the geometric scaling behaviour, being a function of one variable $x^2_{01}\,Q^2_s\Lb Y\Rb$ :
 \beq \label{GS}
   N_{01}\Lb Y,x_{01},b\Rb = N\Lb x^2_{01}\,Q^2_s\Lb Y,b\Rb \Rb
   \eeq
 It is instructive to note, that this behaviour has been proven on general theoretical grounds\cite{BALE}  and has been seen in the experimental data on DIS\cite{GS}.
 
 Finally, in Ref.\cite{LETU} the solution to \eq{BK} was found  deep into saturation region for $z \,\gg\,1$ which has the following form:
  \beq \label{BKS1}
N_{01}\Lb z\Rb\,\,\,=\,\,1\,\,\,-\,\,\,\mbox{Const}\,\exp\Big(  - \frac{z^2}{2\,\kappa}\Big)
 \eeq 
 where ${\rm Const} $ is a smooth function of $z$.

 We have defined the saturation region as $ x^2_{10}\,Q^2_s\Lb Y,b\Rb\,>\,1$. However , in Refs.\cite{GOST,BEST} it has been noted that actually for very large $x_{10} $ the non-linear corrections
 become small
  and we have to solve linear BFKL equation. This feature can be seen  directly
 from the eigenfunction of this equation.  Indeed, the eigenfunction  has the
 following form \cite{LIP}
\beq \label{EIGENF}
\phi_\gamma\Lb \vec{r} , \vec{R}, \vec{b}\Rb\,\,\,=\,\,\,\Lb \frac{
 r^2\,R^2}{\Lb \vec{b}  + \h(\vec{r} - \vec{R})\Rb^2\,\Lb \vec{b} 
 -  \h(\vec{r} - \vec{R})\Rb^2}\Rb^\gamma\,\,\xrightarrow{b\,\gg\,r,rR}\,\,\Lb \frac{ R^2\,r^2}{b^4}\Rb^\gamma\,\,\equiv\,\,e^{\gamma\,\xi}~~\mbox{with}\,\,0 \,<\,Re\gamma\,<\,1\eeq
for any kernel which satisfies the conformal symmetry. In \eq{EIGENF} $R$ is the size of the initial dipole at $Y=0$  while $r\equiv x_{10}$ is the size of the dipole with rapidity $Y$.

One can see that for $r=x_{10} \,>\,min[R, b]$,  $\phi_\gamma$ starts to be small and the non-linear term in the BK equation 
could be neglected.  In this paper we wish to discuss the DIS for $Q^2 \geq 1 \,GeV^2$ and in the region of small $x$. In other words we consider $x_{10} < R$, where $R$ is the radius of a hadron.  Bearing this in mind we can replace \eq{EIGENF} by the following one:
\beq \label{EIGENF1}
\phi_\gamma\Lb \vec{r} , \vec{R}, \vec{b}\Rb\,\,\,=\,\,\,\Lb \frac{
 r^2\,R^2}{\Lb \vec{b}  - \h\vec{R}\Rb^2\,\Lb \vec{b} 
 + \h \vec{R})\Rb^2}\Rb^\gamma  =  \Lb r^2 Q_s\Lb Y=0, b, R\Rb\Rb^\gamma
 \eeq 
 Therefore, we can absorb all impact parameter dependence in the dependence of the saturation scale on $b$.
  
 The main goal  is to find the multiplicity distributions and the entropy of produced gluons in the kinematical region of \eq{BKS1}. In other words, we wish to decipher the saturated amplitude in terms of produced gluons. Fortunately, the equations for the cross section of production of $n$ gluons have been derived in the framework of CGC (see Ref.\cite{KLP}). The next section is a  brief review of the results of Ref.\cite{KLP}. In this section we emphasize the two main ingredients on which the derivation of the equations is based: the AGK cutting rules\cite{AGK} and the BFKL equation\cite{BFKL} , which gives the production of gluons in  a  particular kinematic region typical for  leading log(1/x) approximation of perturbative QCD.
 
 In section 3 we discuss solutions to the equations for  the total cross section $ \sigma_{tot}$, for the cross section of diffraction production  $\sigma_{sd}$ and the cross section for productions of $n$ gluons $\sigma_n$. We show that both the total cross section and the cross section of the diffraction production reaches the unitarity limit: $\sigma_{tot}(b) = 2$ and $\sigma_{sd} = 1$, while $\sigma_n $ are small and proportional to $\,\frac{1}{\bar{n}(z)} \exp\Lb - n/\bar{n}(z)\Rb$ with $\bar{n}(z) \,\,\propto\,\, \exp\Big( \frac{z^2}{2\,\kappa}\Big)$. In section 4 we estimate the value of the entropy of produced gluons. In conclusions we summarize our results and discussed related problems.

 
  \section{  Generating functional for multipartical production processes } 
 
 
  \subsection{  Generating functional for the scattering amplitude } 
\begin{figure}[ht]
    \centering
  \leavevmode
      \includegraphics[width=8cm]{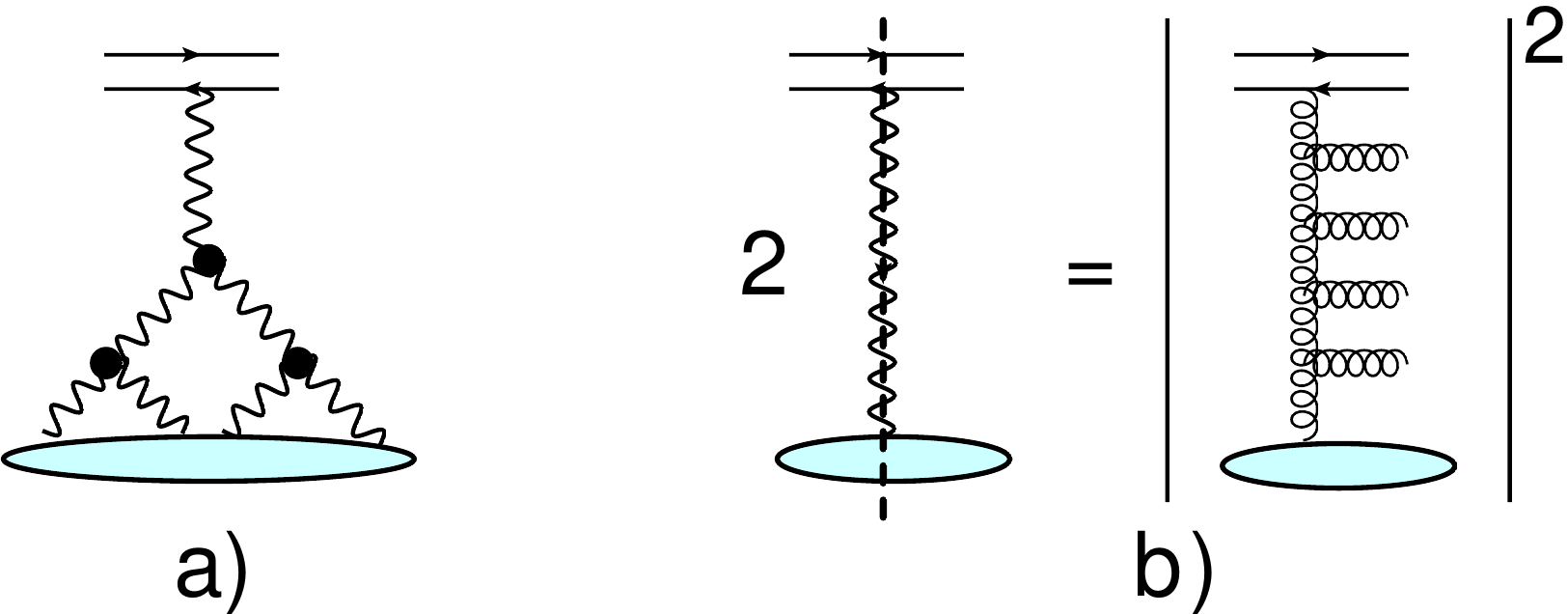}  
      \caption{ \fig{fan}-a:The  fan diagrams in the BFKL Pomeron calculus that describe the BK equation (see \eq{BK}).\fig{fan}-b: The unitarity constraints for the exchange of the BFKL Pomeron: the structure of the cut Pomeron. This constraints  determines the cut Pomerons that  are shown in \fig{agk} by the wavvy lines which are crossed by the dashed ones.}
\label{fan}
   \end{figure}

 The scattering amplitude, given by BK equation (see \eq{BK}) , can be viewed as a sum of the "fan" BFKL Pomeron diagrams\cite{GLR,MUQI,BART,BRN,BRAUN,KOLU}(see \fig{fan}). The value of the triple Pomeron vertex is determined by the non-linear term of the BK equation.  However, it has been shown in Ref/\cite{MUDI} that the BK equation has a quite different interpretation in which the BFKL kernel 
 
  \beq \label{K}
\frac{ \bas}{2\,\pi} K\Lb x_{01}|x_{02},x_{12}\Rb \,\,=\,\,\frac{ \bas}{2\,\pi} \frac{ x^2_{01}}{x^2_{02}\,x^2_{12}} 
 \eeq
 gives the probability for decay of one dipole with size $x_{10}$ to two dipoles with sizes  $x_{02}$  and $x_{12}$.
 
The simplest and the most transparent technique to incorporate this decay is the generating functional which allows us to reduce the calculation of the high energy elastic amplitude to consideration of a Markov process.   
The  generating functional is defined as \cite{MUDI,MUSA,Salam,LELU}
\beq \label{Z11}
Z_0\left(Y;\{u\} \right)\,\,\equiv\,\,\sum_{n=1}\,\int\,\,
P_n\left(Y; r_1, \dots  ,r_n \right) \,\,\prod^{n}_{i=1}\,u(r_i) 
d^2r_i
\eeq 
where $u(r_i)$ is an arbitrary function of $r_i$ and $b_i$. $P_n$ is the probability density to 
find  $n$ dipoles with sizes $r_1,\dots,r_n$ at  rapidity $Y$. 
 \footnote{The dipole $(x_i, y_i)$ with coordinates $x_i$ for quark  and $y_i$ for antiquark can be characterized by the dipole size $\vec{r}_i = \vec{x}_i  - \vec{y}_i$ and $ \vec{b}_i = \h ( \vec{x}_i  + \vec{y}_i)$ . For simplicity we suppress in \eq{Z11} and below the coordinate $b_i$. For the scattering with the nuclear target we can consider  that impact parameters of all dipoles are the same $b_i = b$ (see Ref. \cite{BK}). The alternative notations are $\vec{r}_i =
  \vec{r}_{ik} = |\vec{x}_i - \vec{y}_{k = i+1} |$.} 
 For functional of \eq{Z11} we have two obvious  conditions:
 \begin{subequations} 
    \bea 
 \mbox{initial conditions:}  & &Y =0 \,\, P_n\,=\,0\,\, \mbox{for\,\,\,}  n \,>\,1\,\, \mbox{and}\,\, P_1 = \delta( \vec{r} 
\,-\,\vec{r}_1)\delta(\vec{b} - \vec{b}_1);Z_0\left(Y =0;\{u\}\right)\,\,=\,\,u(r)\,\,;\label{IC}\\ 
 \mbox{boundary conditions:}  & & u =1\,\,
Z_0\left(Y;\{u\}\right)|_{u=1}\,\,=\,\,1;\label{BC}
\eea
  \end{subequations} 
\eq{BC} follows from the physical meaning of $P_n$ and represents the conservation of the total probability, while \eq{IC} indicates that we are considering the interaction of one dipole with the target.

 The Markov process can be described as the following equation for the generating functional:
 \beq \label{Z3}
\frac{d Z_0\left(Y;\{u\} \right)}{d
Y}= 
  \,\frac{\bas}{2 \pi}\,\int\,\,d^2 r \;d^2 r' \,K\Lb r | r', |\vec{r} - \vec{r'}|\Rb \left\{\,u(r)\,
-\,
u(r')\,u(|\vec{r'} - \vec{r}|)\right\}
\,\frac{\delta}{\delta u(r)} \,\, Z_0\left(Y -y;\{ u\} \right)
\eeq
 Two terms of this equation has simple meaning of 
increase of  the probability to find $n$-dipoles due to decay of one  dipole to two dipoles (birth terms, the second term in \eq{Z3}) and of decrease of the probability since one of $n$-dipoles can decay (death term, the first term in \eq{Z3})
    
 \eq{Z3} is general and can be used for arbitrary initial condition. In the case of \eq{IC} it can be rewritten as the non-linear equation.
 \eq{Z3} being a linear equation with only first derivative has a general solution of the form 
$Z_0\left( Y-y;\{u\}\right) \equiv 
Z_0\left(\{u(Y-y)\} \right)$. Inserting this solution and using the initial condition of \eq{IC} we obtain the non-linear equation
\beq  \label{Z4}
\frac{d Z_0\left(Y;\{u\}\right)}{d  Y}=\frac{\bas}{2 \pi}\int d^2 r'
 \,K\Lb r| r', | \vec{r} - \vec{r'}|\Rb \left\{Z_0\left(Y,\{u\}\right) -
Z^2_0\left(Y,\{u\}\right)\right\}
\eeq

It is easy to see that \eq{Z4} can be re-written as the Balitsky-Kovchegov  equation \cite{BK} for the scattering amplitude (see for example Ref.\cite{KOLEB}).   
 
 
  \subsection{ AGK cutting rules and  generating functional for the production of gluons } 
\begin{figure}[ht]
    \centering
  \leavevmode
      \includegraphics[width=14cm]{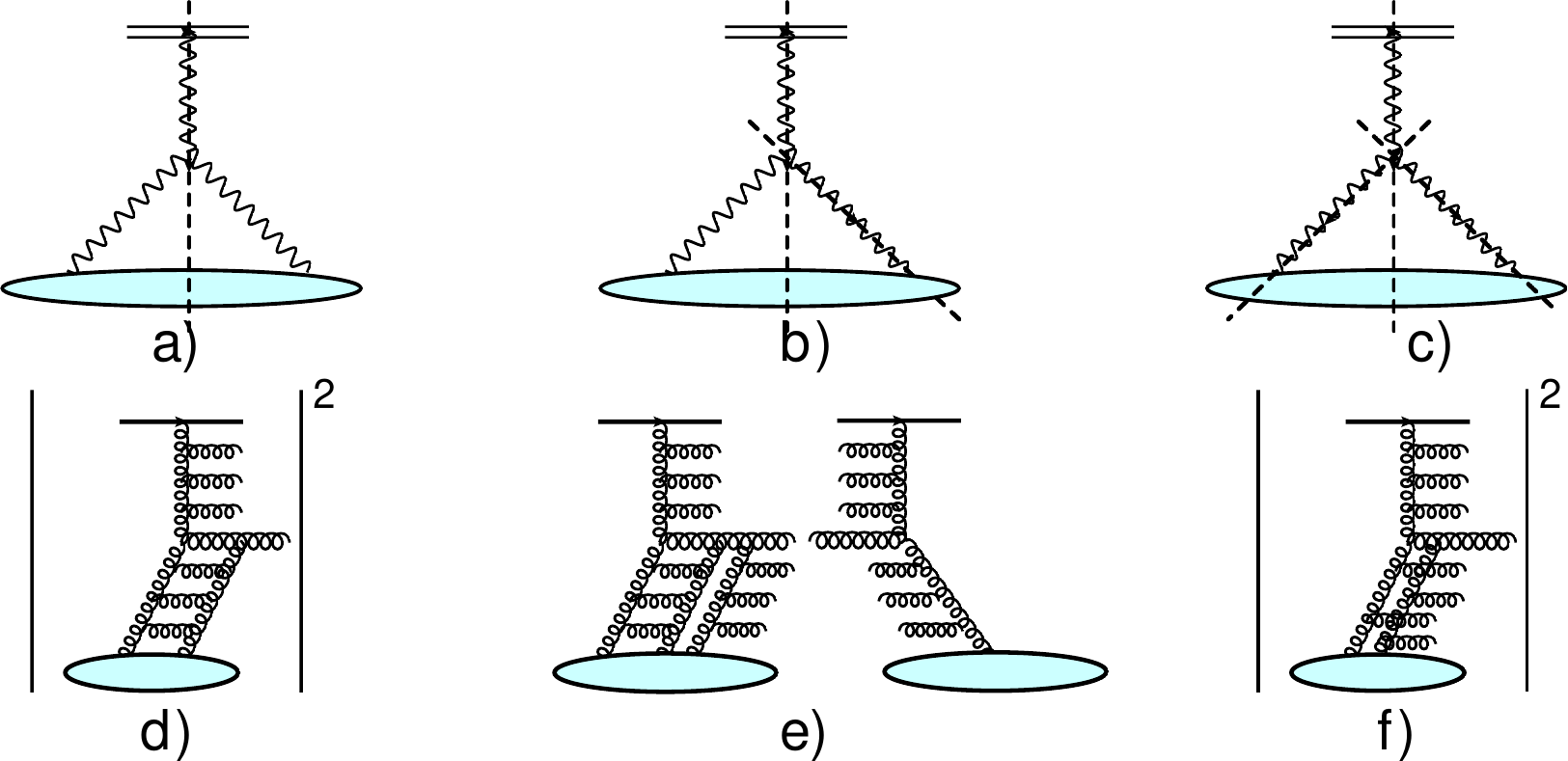}  
      \caption{ Three vertices  for  gluon production accordingly to AGK cutting rules for  fan diagrams in the BFKL Pomeron calculus.  The wavy lines denote the BFKL Pomerons.  The wavy lines crossed by the dashed ones show the cut  Pomerons.}
\label{agk}
   \end{figure}

 Based on this probability interpretation of the sum of "fan'' diagrams in Ref.\cite{KLP} an attempt was made to develop the probability approach for the production of the gluon using the AGK   cutting rules\cite{AGK}. It is worthwhile mentioning that such kind of approach was successfully applied for interactions of two BFKL cascades in the simple case with zero transverse dimension \cite{MUSA, Salam,LEPRSM}. The AGK cutting rules use that the BFKL Pomeron\cite{BFKL} gives the cross section of produced gluons in the specific kinematic region  typical for leading log(1/x) approximation of perturbative QCD. 
These cutting rules allow us to expand the contribution of the exchange of $n$ BFKL Pomerons  as a definite sum of the final states with the average  number of produced gluon $ k \Delta Y$, where $k\, \leq\, n$ and $\Delta \,Y $ is the mean multiplicity for one BFKL Pomeron exchange. Actually, the AGK cutting rules introduce three different triple Pomeron vertices (see \fig{agk} ). 
 The structure of the production processes is determined by the s-channel unitarity constraints for the BFKL Pomeron:
 \beq \label{UNPO}
2 N^{BFKL}\Lb Y; r, b \Rb \,\,=\,\, G^{BFKL}_{in}\Lb Y;r,b\Rb
\eeq  
   where  $N^{BFKL}\Lb Y; r, b \Rb$ is the contribution to the scattering amplitude by the BFKL Pomeron $ G^{BFKL}_{in}\Lb Y;r,b\Rb$ is the cross section of produced gluons, which was actually calculated in Ref.\cite{BFKL} and which is shown in \fig{fan}-b. In framework of the AGK cutting rules it  is  called the cut Pomeron which is shown in \fig{agk} by the wavy lines which are crossed by the dashed ones.
   
   In Ref.\cite{KLP}  it is    introduced a generalization of \eq{Z11}:
\bea \label{Zmulti}
 Z \left(Y;\{u\} ,\{v\}\right)\,\,\equiv 
  \,\,\sum^{\infty}_{n=0,m=0}\,\int\,\,
P^m_n\left(y; r_1, \dots ,r_n; r_1,\dots , r_m \right) \,\,\prod^{n}_{i=1}\,u(r_i) \prod^{m}_{k=1}\,v(r_k) 
d^2r_i\,d^2r_k 
\eea
where $P^m_n$ is the probability to find  (i) $n$ dipoles in the wave function of the fast dipole, which interact with the target at time  $\tau =0$; and (ii) $m$ dipoles which can be detected at $\tau=\infty$.  In other words,        $P^m_n$ is  the probability to have $n$   dipoles with sizes $r_1, \dots, r_n$ at $\tau = 0 $  at rapidity $Y$, which do not survive until $\tau=\infty$ and cannot be measured, while the dipoles with sizes $
r_1, \dots, r_m$  reach $\tau = \infty$ and can be caught by detectors.

The initial conditions of \eq{IC} has to be replaced by 
\beq \label{ICMP}
Z\left(Y =0;\{u\}, \{v\}\right)\,\,=\,\,v(r)    
    \eeq
    and we have two boundary conditions:
   \beq \label{BCMP} 
    (1)~~Z \left(Y;\{u\} ,\{v\}\right)|_{u=1,v=1} \,\,=\,\,1; ~~~~~(2) ~~Z\Lb Y; \{u\},\{ v \}\Rb |_{v=2u-1} =2 \;Z_0\Lb Y; \{u \}\Rb\,\,-1;
    \eeq
   The first one comes from the conservation of probabilities while the second stems from the s-channel unitarity:
   \beq \label{UNIT}
   2 N\Lb Y, r, b\Rb\,\,=\,\,\sigma_{sd}\Lb Y, r,b\Rb \,\,+\,\,\sigma_{in} \Lb Y, r, b\Rb
   \eeq
   where $\sigma_{sd}$ and $\sigma_{in}$ are the single diffraction and inelastic cross sections at fixed impact parameter $b$.  
   
   In Ref.\cite{KLP} the AGK cutting rules for the ''fan'' diagrams  were reduced to the following linear equation for 
 $  Z \left(Y;\{u\} ,\{\zeta = 2\,u -  v\}\right)  $: 
\bea \label{MPZLEQ}
&&\frac{\partial \tilde{Z} \left(Y;\{u\} ,\{\zeta \}\right)}{\partial Y}\,\,=\,\,\frac{\bas}{2 \pi}\,\int\,d^2 r_2 \,K\Lb r_{10}| r_{12}, r_{02}\Rb \times\\
&&\left\{ \Lb u(r_{12})\,u(r_{02})\,-\,u(r_{10}) \Rb \frac{\delta  \tilde{Z} \left(y;\{u\} ,\{\zeta\}\right)}{\delta  u(r_{10})}\,\,+\,\,\Lb \zeta(r_{12})\,\zeta(r_{02})\,-\,\zeta(r_{10}) \Rb\,\frac{\delta  \tilde{Z} \left(y;\{u\} ,\{\zeta \}\right)}{\delta  \zeta(r_{10})}\right\}\nn
\eea
Each term in \eq{MPZLEQ} has the same transparent probabilistic interpretation as in \eq{Z3}.
 \begin{boldmath}
  \subsection{ The master equation for $\sigma_n$}  
  \end{boldmath}
   \eq{MPZLEQ} can be used for an arbitrary initial condition. For  DIS scattering the initial condition of \eq{ICMP}
   results in the non-linear equation. This equation takes the most elegant form for the generating functional
    which is defined   as
\bea
&&M\Lb Y-Y_0; r,b\Rb \,\, = \,\,  1\,\,\,-\,\, Z\Lb Y-Y_0; \{u\},\{ v\}  \Rb\label{MP}\\
&&
=\,\sum^{\infty}_{n=1,m=0}\,\,\frac{ 
(-1)^{n + m + 1}}{n!\,m!}\,\,\left\{\int\,\prod^n_{i =1}\,\,d^2r_i\,\int\,\prod^m_{k =1}\,\,d^2r_k\,\gamma(r_i)\,\gamma_{in}(r_k) \,\,
\frac{\delta}{\delta u(r_i) } \,\,\frac{\delta}{\delta v(r_k) }\right\} \,Z\left(Y-Y_0); \{u\} ,\{v\}\right)\Big{|}_{u(r)=1, v(r)=1}\nn
\eea   
   where $u(r) = 1 - \gamma(r)$ and $v(r) = 1 - \gamma_{in}(r)$.
   
   For $M\Lb Y; r,b\Rb$  the non-linear equation  is obtained from \eq{MPZLEQ} in Ref.\cite{KLP}:
  \bea \label{MPNEQ}
&& \frac{\partial M(Y;r_{10},b)}{\partial Y}\,=\,
\,\,\frac{\bas}{2 \pi}\,\int\,d^2 r_2 \,K\Lb r_{10}| r_{12}, r_{02}\Rb \left\{M(Y;r_{12},b)\,+\,\begin{scriptsize}\begin{footnotesize}\end{footnotesize}\end{scriptsize}
M(Y;r_{20},b)\,-
\,M(Y;r_{10},b)\,\right.  \\
&& \left.+
M(Y;r_{12},b)M(Y;r_{20},b)
\,-\,2\,M(Y;r_{12},b)N(Y;r_{20},b) -\,2\,
N(Y;r_{12},b)
M(Y;r_{20},b)
\,+\, 2
N(Y;r_{12},b)
N(Y;r_{20},b)
\right\}\notag
 \eea   
   The first check of this equation: $ M(y;r_{10},b   ) $ for $\gamma_{in}=0$ gives the cross section of diffraction production $M(y;r_{10},b   )|_{\gamma_{in}=0}  = \sigma_{sd}$.  This equation coincides with the equation for the  diffraction production of Ref.\cite{KOLE}.
   
  If we want to find a cross section with the $k$-produced gluons in dipole nucleus (or proton)  interaction we need to calculate
\beq \label{XSK}
\sigma_k\Lb Y,r; b \Rb \,\,\,=\,\,\frac{1}{k!}\,\prod^k_{i = 1}\,\gamma_{in}(r_i)\left(\frac{\delta}{\delta \gamma_{in}(r_i)}\,M \left(Y;\{\gamma\} ,\{\gamma_{in} \}\right)\right)\Big{|}_{\gamma_{in}=0}
\eeq
where $\gamma(r)$ is the low energy elastic amplitude and $\gamma_{in}(r) = 2 \gamma(r)$ at low energy ($Y_0$).

~


  \subsection{Solutions for Pomeron calculus  in zero transverse dimension}  

 For the Pomeron calculus in zero transverse dimension \eq{MPZLEQ} and \eq{MPNEQ} were solved in Ref.\cite{LEPRSM}. The solutions  for the generating functions take the forms:
  \begin{subequations} 
   \bea
 Z_0\Lb Y, u\Rb &=&\,\frac{u e^{ - \Delta \,Y}}{ 1 - u \Lb 1 \,-\,e^{ - \Delta \,Y} \Rb}; \label{ZSM1}\\
 Z\Lb Y, u, v \Rb &=& \frac{2u e^{ - \Delta \,Y}}{ 1 - u \Lb 1 \,-\,e^{ - \Delta \,Y} \Rb} \,-\,\frac{(2u -v) e^{ - \Delta \,Y}}{ 1 - (2u -v)\Lb 1 \,-\,e^{ - \Delta \,Y} \Rb}; \label{ ZSM2}
   \eea
   \end{subequations}  
 where $\Delta$ is the intercept of the BFKL Pomeron. From \eq{XSK} we obtain:
 
 \beq\label{XSKSM}
 \sigma_k\,\,=\,\,\frac{e^{ \Delta \,Y }\,\Lb e^{  \Delta \,Y}\,-\,1 \Rb^{k-1}}{\Lb1 - 2\,\gamma \Lb e^{  \Delta \,Y}\,-\,1 \Rb\Rb^{k+1}}\,\gamma_{in}^k 
 \eeq
 where $\gamma$  and $\gamma_{in}$ are  the scattering amplitude  and [production cross section of two dipoles at low energy. From \eq{UNPO} $\gamma_{in} = 2 \gamma$.
 
 Using \eq{ZSM1} we can calculate the scattering amplitude, which has the following form:
 \beq \label{NSM}
 N\Lb Y\Rb\,\,=\,\,\frac{\gamma}{1 - (1 - \gamma)\Lb 1\,\,-\,\,e^{ -\Delta \,Y }\Rb }\,\,=\,\,
 \frac{\tilde{\gamma}  e^{ \Delta \,Y } }{\tilde{\gamma} e^{ \Delta \,Y }\,  -\,1}
 \eeq
 where $\tilde{\gamma} = \gamma/(1 - \gamma)$.
 
 Considering $e^{ \Delta \,Y } $ as  Pomeron Green's function and using the AGK cutting rules for the scattering amplitude of \eq{NSM} we reproduce at large values of $Y$  the same $\sigma_k$ which follow from \eq{XSKSM}. We believe that this
 observation is a strong argument for  that our probabilistic treatment of gluon production is correct.

 
  \section{ The cross sections of gluon production deep in the saturation region } 
The main goal of this paper is to find  $\sigma_k$ solving \eq{MPNEQ} in the kinematic region where we have 
the scattering amplitude  of \eq{BKS1}. In other words, we are looking for the solutions for the scattering anplitude of the dipole with size $r \equiv r_{01}$ and rapidity $Y$ in the kinematic region where $r^2 Q^2_s(Y, b)\,\,\gg\,\,1$.

 
  \subsection{ The total cross section } 
 The total cross section can be obtained  considering $M\Lb Y, r, b\Rb$ for $\gamma_{in}(r_i) = 2 \gamma(r_i)$. 
 From the unitarity constraints of \eq{UNIT} we expect that in our kinematic region $\sigma_{tot}\Lb Y, r,b\Rb$ approaches  the unitarity limit of $\sigma_{tot} \to 2$. Introducing $\sigma_{tot}\Lb Y, r,b\Rb\,\,=\,\,2\,\,-\,\,\Delta_{tot}\Lb Y, r,b\Rb$ and $N \Lb Y, r,b\Rb \,\,=\,\,1\,\,-\,\,\Delta_0\Lb Y, r, b\Rb$ , we can rewrite \eq{MPNEQ}  in the following form. neglecting the contribution of $\Delta_{in}^2$ or/and $\Delta_0^2$:
 \bea \label{INXS1}
 \frac{\partial \Delta_{tot} (Y;r_{01},b)}{\partial Y} \,\,&=&\,\,
\frac{\bas}{2 \pi} \!\!\!\!\! \int\limits_
 {\begin{subarray}{l}
 r_{02} \,\gg\,1/Q_s(Y)\\
r_{12}\,\gg\,1/Q_s(Y)
\end{subarray}}
\!\!\!\!\!\!\!\!\!\!\!\!\!d^2 r_2 \,K\Lb r_{01}| r_{12}, r_{02}\Rb 
\Lb 
  \Delta_{tot} (Y;r_{02},b) \,\,+\,\, \Delta_{tot} (Y;r_{12},b)\,\, -\, \,\Delta_{tot} (Y;r_{01},b)\Rb\nn\\
 &-\,&2\,\frac{\bas}{2 \pi}\!\!\!\!\! \int\limits_
 {\begin{subarray}{l}
 r_{02} \,\gg\,1/Q_s(Y)\\
r_{12}\,\gg\,1/Q_s(Y)
\end{subarray}}
\!\!\!\!\!\!\!\!\!\!\!\!\!d^2 r_2\,K\Lb r_{01}| r_{12}, r_{02}\Rb \Lb 
  \Delta_{0} (Y;r_{02},b) ,\,+\,\, \Delta_{0} (Y;r_{12},b) \Rb
  \eea
 
 One can see that \eq{INXS1}  has  solution: $ \Delta_{tot} (Y;r_{01},b)\,=\,2\, \Delta_{0} (Y;r_{01},b)$. Indeed, substituting   this solution into \eq{INXS1}  we obtain the equation for $\Delta_{0} (Y;r_{01},b)$:
 \beq \label{INXS2}
 \frac{\partial \Delta_{0} (Y;r_{01},b)}{\partial Y} \,\,=\,\,
-\,\Lb \frac{\bas}{2 \pi} \!\!\!\!\! \int\limits_
 {\begin{subarray}{l}
 r_{02} \,\gg\,1/Q_s(Y)\\
r_{12}\,\gg\,1/Q_s(Y)
\end{subarray}}
\!\!\!\!\!\!\!\!\!\!\!\!\!d^2 r_2 \,K\Lb r_{01}| r_{12}, r_{02}\Rb \Rb
\,\Delta_{0} (Y;r_{01},b)\,\,=\,\,- z\,\,\Delta_{0} (Y;r_{01},b)
\eeq
which is the equation for $\Delta_0$, that has been discussed in Ref.\cite{LETU}, and from which follows the asymptotic behaviour of \eq{BKS1}. 
 
 
  \subsection{ The  diffraction production } 
  As we have discussed the master equation (see \eq{MPNEQ}) coincides with the equation of Ref.\cite{KOLE}, which has been derived  in a  quite different  way than it is done here. For the cross section of diffraction production we need to consider   $M(Y;r_{10},b) $ of \eq{MP} at $\gamma_{in}\Lb r\Rb=0$.  From \eq{UNIT} we expect that $\sigma_{in}(Y; r_{10},b)$ approaches 1 at high energies. Bearing this in mind we 
 introduce $\sigma_{sd}\Lb Y, r,b\Rb\,\,=\,\,1\,\,-\,\,\Delta_{sd}\Lb Y, r,b\Rb$ and $N \Lb Y, r,b\Rb \,\,=\,\,1\,\,-\,\,\Delta_0\Lb Y, r, b\Rb$. Plugging these expressions in \eq{MPNEQ}  we see that
 \beq \label{SD1}
 \frac{\partial \Delta_{sd} (Y;r_{01},b)}{\partial Y} \,\,=\,\,
-\,\Lb \frac{\bas}{2 \pi} \!\!\!\!\! \int\limits_
 {\begin{subarray}{l}
 r_{02} \,\gg\,1/Q_s(Y)\\
r_{12}\,\gg\,1/Q_s(Y)
\end{subarray}}
\!\!\!\!\!\!\!\!\!\!\!\!\!d^2 r_2 \,K\Lb r_{01}| r_{12}, r_{02}\Rb \Rb
\,\Delta_{sd} (Y;r_{01},b)\,\,=\,\,- z\,\,\Delta_{sd} (Y;r_{01},b).
\eeq
  In the variable $z$ of \eq{z} \eq{SD1} takes the form:
  
  \beq \label{SD2}
\kappa \frac{d\,  \Delta_{sd} (z)}{d \,z}  \,\,=\,\,- z\,\, \Delta_{sd} (z)
\eeq
with the solution:
 \beq \label{SD2}
 \Delta_{sd} (z)  \,\,=\,\,C(z)\exp\Lb - \frac{z^2}{ 2\,\kappa}\Rb
\eeq  
  where $\kappa$ is given by \eq{GACR} and $C(z)$ is a smooth function of $z$.  Therefore, $\sigma_{sd}\Lb Y, r,b\Rb$ at large values  of $Y$ has the form of \eq{BKS1}  and shows the geometric scaling behaviour as it has been discussed in Ref.\cite{KOLE}.

  It is instructive to note that  $\sigma_{tot}\Lb Y, r,b\Rb\,\,=\,\,2\,\,-\,\,\Delta_{in}\Lb Y, r,b\Rb$ and  $\sigma_{sd}\Lb Y, r,b\Rb\,\,=\,\,1\,\,-\,\,\Delta_{sd}\Lb Y, r,b\Rb$    are the solutions of the same equation. We believe that this observation is the important check that the main ideas of Ref.\cite{KLP}, which result in deriving \eq{MPNEQ}, are correct.

 \begin{boldmath}
  \subsection{ $\sigma_1\Lb Y, r,b\Rb$} 
  \end{boldmath}
   
   Using \eq{XSK} 
   we can obtain the equations for $\sigma_k$ from the master equation by differentiating this equation. First taking 
   $\frac{\delta}{\delta \gamma_{in}(r_i)  }$ from both parts of \eq{MPNEQ} and putting  $ \gamma_{in}(r_i)  = 0$   we obtain the following equation for $\sigma_1\Lb Y, r,b\Rb$:
   \bea \label{XS1}
   \frac{\partial \sigma_1 (Y;r_{01},b)}{\partial Y} \,\,&=&\,\,
 \frac{\bas}{2 \pi} \!\!\!\!\! \int\limits_
 {\begin{subarray}{l}
 r_{02} \,\gg\,1/Q_s(Y)\\
r_{12}\,\gg\,1/Q_s(Y)
\end{subarray}}
\!\!\!\!\!\!\!\!\!\!\!\!\!d^2 r_2 \,K\Lb r_{01}| r_{12}, r_{02}\Rb \Bigg\{ \sigma_1\Lb Y,r_{02},b\Rb \,+\,\sigma_1\Lb Y,r_{12},b\Rb\,-\,\sigma_1\Lb Y,r_{01},b\Rb\,\,\nn\\
&+&\,\,\sigma_1\Lb Y,r_{02},b\Rb\,\sigma_{sd} \Lb Y,r_{02},b\Rb\,
+\,
\sigma_1\Lb Y,r_{12},b\Rb\,\sigma_{sd}\Lb Y,r_{02},b\Rb\nn\\
 \,&-&\,2\,\sigma_1\Lb Y,r_{02},b\Rb\,N \Lb Y,r_{02},b\Rb\,-\,2\,
\sigma_1\Lb Y,r_{12},b\Rb\,N\Lb Y,r_{02},b\Rb\Bigg\}
\eea
   
The high energy asymptotic behaviour  we get from \eq{XS1} substituting  $\sigma_{sd}\Lb Y,r_{i,i+1},b\Rb \,=\,1 - \Delta_{sd}\Lb z\Rb$ and $N \Lb Y, r,b\Rb \,\,=\,\,1\,\,-\,\,\Delta_0\Lb Y, r, b\Rb$, \eq{XS1} takes the following form:
  \bea \label{XS2}
   \frac{\partial \sigma_1 (Y;r_{01},b)}{\partial Y} \,\,&=&\,\,-\,
\Lb \frac{\bas}{2 \pi} \!\!\!\!\! \int\limits_
 {\begin{subarray}{l}
 r_{02} \,\gg\,1/Q_s(Y)\\
r_{12}\,\gg\,1/Q_s(Y)
\end{subarray}}
\!\!\!\!\!\!\!\!\!\!\!\!\!d^2 r_2 \,K\Lb r_{01}| r_{12}, r_{02}\Rb\Rb \Bigg\{ \,\sigma_1\Lb Y,r_{01},b\Rb\,-\,\,\sigma_1\Lb Y,r_{02},b\Rb\,\Delta_{in} \Lb Y,r_{02},b\Rb\,\nn\\
& &-\,\,
\sigma_1\Lb Y,r_{12},b\Rb\,\Delta_{in}\Lb Y,r_{02},b\Rb\Bigg\}
\,=\,\,- z\,\sigma_1\Lb Y,r_{01},b\Rb\eea
where $ \Delta_{in}\Lb Y,r_{0i},b\Rb =  2 \Delta_{0}\Lb Y,r_{0i},b\Rb \,-\,\Delta_{sd}\Lb Y,r_{0i},b\Rb$. The total inelastic cross section has the form: $ \sigma_{in} =  1 -  \Delta_{in}\Lb Y,r_{0i},b\Rb $ and  the above equation for $\Delta_{in}$ follows from the unitarity constraints (see \eq{UNIT}).

  Neglecting the second  and the third terms in the curly bracket  one can see that the geometric scaling solution of \eq{XS2} has the same form as \eq{SD2} for $
 \Delta_{sd} (z) $. Hence
 \beq \label{XS3}
\kappa \frac{d\,  \sigma_{1} (z)}{d \,z}  \,\,=\,\,- z\,\, \sigma_{1} (z)~~~\mbox{with the solution}~~~
 \sigma_1 (z)  \,\,=\,\,C(z)\exp\Lb - \frac{z^2}{ 2\,\kappa}\Rb
\eeq  
We will discuss below the  influence of the neglected  terms on the solution of \eq{XS3}

It turns out that it is convenient to discuss the solutions in the momentum representation, where \eq{XS3} takes the form of convolution.
  Adding and subtracting the gluon reggeization term in the momentum representation we obtain the following equation for the geometric scaling solution:

\beq \label{XS1MR}
\kappa\frac{d \,  \sigma_{1} (\zz)}{d \,\zz} \,\,=\,\,\,\, \int \frac{d^2 k'_T}{(2 \pi)^2} K\Lb \vec{k}_T,\vec{k}'_T\Rb\,\,\sigma_1\Lb \zz' \Rb\,\,\,-\,\,\zz\,\sigma_1\Lb \zz\Rb\,\,+\,\, \Delta_{in}\Lb \zz\Rb \sigma_1\Lb \zz\Rb \eeq

where  $K\Lb \vec{k}_T,\vec{k}'_T\Rb$ is the BFKL kernel in momentum representation:

\beq \label{KERM}
K\Lb \vec{k}_T,\vec{k}'_T\Rb\,\,\sigma_1\Lb Y, \vec{k}'_T, b \Rb\,=\,\frac{1}{\Lb \vec{k}_T - \vec{k}'_T\Rb^2} \,\,\,\sigma_1\Lb Y, \vec{k}'_T, b \Rb\,\,-\,\,\frac{k^2_T}{\Lb \vec{k}_T - \vec{k}'_T\Rb^2\,\Lb\Lb \vec{k}_T - \vec{k}'_T\Rb^2\,+\,k'^2_T\Rb }\,\,\sigma_1\Lb Y, \vec{k}_T, b \Rb\eeq

$\zz$ is a new scaling variable:
\beq \label{zm}
\zz\,\,\,=\,\,\kappa\,\bas \,Y \,\,-\,\,\ln k^2_T\,\,=\,\,\ln\Lb \frac{Q^2_s\Lb Y, b\Rb}{k^2_T}\Rb
\eeq
Using the Mellin transform:
\beq \label{MET}
\sigma_1\Lb\zz, b \Rb\,\,=\,\,\int\limits^{\epsilon + i \infty}_{\epsilon - i \infty} \frac{d \gamma}{2\,\pi\,i} e^{ \gamma\,\zz} \,\sigma_1\Lb \gamma, b\Rb
\eeq
we can rewrite \eq{XS1MR} in the form:
\beq \label{XS1MR1}
\Lb \kappa\,\gamma\,\,-\,\,\chi\Lb \gamma\Rb\Rb \sigma_1\Lb \gamma, b\Rb =\frac{ d \sigma_1\Lb \gamma, b\Rb}{d\,\,\gamma}
  \eeq
  
  with the solution
  \beq \label{XS1MR2}  
  \sigma_1\Lb \gamma, b\Rb\,\,=\,\,\sigma^{init}_{1} \Lb b\Rb \exp\Lb  \h \kappa \,\gamma^2 \,\,+\,\int^\gamma_{\h} \chi\Lb \gamma'\Rb d \gamma'\Rb
  \eeq  
  where $\sigma^{init}_1$ has to be found from the initial conditions. Plugging \eq{XS1MR2} into  \eq{MET} we can take the integral over $\gamma$ using the method of steepest decent. The equation for the saddle point takes the form:
 \beq \label{XS1MR3}  
\zz + \kappa \gamma_{SP} \,+\, \chi\Lb   \gamma_{SP}\Rb=0
\eeq
Since $\chi\Lb \gamma_{SP}\Rb \sim \ln \gamma_{SP}$  the value of $\gamma_{SP} = -\zz/\kappa$.  Hence
  \beq \label{XS1MR4}  
\sigma_1\Lb \zz\Rb\,\,=\,\,C\Lb \zz\Rb e^{ - \frac{\zz^2}{2\,\kappa}}
\eeq
 with $C\Lb \zz\Rb$  being a smooth function of $\zz$. Therefore, at large values of $Y$ the solutions in coordinate and momentum representations have the same form in $z$ and $\zz$, respectively. The last term in \eq{XS1MR} leads to replacement of $\zz  \to \zz  -  \intl^{\infty}_0 d\zz' \,\Delta_{in}\Lb \zz\Rb$ in \eq{XS1MR4}. One can see this solving the simplified equation:
 \beq \label{XS1MR5}
\kappa\frac{d \,  \sigma_{1} (\zz)}{d \,\zz} \,\,=\,\,\,\, \,\,\,-\,\,\zz\,\sigma_1\Lb \zz\Rb\,\,+\,\, \Delta_{in}\Lb \zz\Rb \sigma_1\Lb \zz\Rb
\eeq
with the solution:
\bea \label{XS1MR6}
&&\sigma_1\Lb \zz\Rb =  {\rm C} \exp\Lb  - \frac{\zz^2}{2\,\kappa}  + \frac{1}{\kappa}\int^{\zz}_0 d \zz'  \,\Delta_{in}\Lb \zz'\Rb \Rb \nn\\
&&  =  {\rm C} \exp\Lb  - \frac{\zz^2}{2\,\kappa}  + \frac{1}{\kappa} \int^{\infty}_0 d \zz' 
\,\Delta_{in}\Lb \zz'\Rb   -  
\underbrace{ \frac{1}{\kappa}\int^{\infty}_{\zz}  d \zz'  \,\Delta_{in}\Lb \zz'\Rb}_{\ll 1}\Rb= {\rm C'}\exp\Lb - \frac{\Lb  \zz - \int^{\infty}_{\zz}  d \zz'  \,\Delta_{in}\Lb \zz'\Rb\Rb^2}{2\,\kappa}\Rb   \eea

  ~

 \begin{boldmath}
  \subsection{ $\sigma_2\Lb Y, r,b\Rb$} 
  \end{boldmath}
       
    Applying $\h\frac{\delta}{\delta \gamma_{in}(r_1)}\,\frac{\delta}{\delta \gamma_{in}(r_2)  }$ to the both parts of the master equation we obtain:
     \bea \label{XS20}
   \frac{\partial \sigma_2 (Y;r_{01},b)}{\partial Y} \,\,&=&\,\,
 \frac{\bas}{2 \pi} \!\!\!\!\! \int\limits_
 {\begin{subarray}{l}
 r_{02} \,\gg\,1/Q_s(Y)\\
r_{12}\,\gg\,1/Q_s(Y)
\end{subarray}}
\!\!\!\!\!\!\!\!\!\!\!\!\!d^2 r_2 \,K\Lb r_{01}| r_{12}, r_{02}\Rb \Bigg\{ \sigma_2\Lb Y,r_{02},b\Rb \,+\,\sigma_2\Lb Y,r_{12},b\Rb\,-\,\sigma_2\Lb Y,r_{01},b\Rb\,\,\nn\\
&+&\,\,\sigma_2\Lb Y,r_{02},b\Rb\,\sigma_{sd} \Lb Y,r_{02},b\Rb\,
+\,\sigma_1\Lb Y,r_{12},b\Rb\,\sigma_{1}\Lb Y,r_{02},b\Rb\,+\,\sigma_2\Lb Y,r_{12},b\Rb\,\sigma_{sd}\Lb Y,r_{02},b\Rb\nn\\
 \,&-&\,2\,\sigma_2\Lb Y,r_{02},b\Rb\,N \Lb Y,r_{02},b\Rb\,-\,2\,
\sigma_2\Lb Y,r_{12},b\Rb\,N\Lb Y,r_{02},b\Rb\Bigg\}
\eea
  Replacing at large $Y$ $\sigma_{sd}\Lb Y, r,b\Rb\,\,=\,\,1\,\,-\,\,\Delta_{sd}\Lb Y, r,b\Rb$ and $N \Lb Y, r,b\Rb \,\,=\,\,1\,\,-\,\,\Delta_0\Lb Y, r, b\Rb$
  we  reduce \eq{XS2}   to the form: 
       \bea \label{XS21}
   \frac{\partial \sigma_2 (Y;r_{01},b)}{\partial Y} \,\,&=&\,\,-\,
\Lb \frac{\bas}{2 \pi} \!\!\!\!\! \int\limits_
 {\begin{subarray}{l}
 r_{02} \,\gg\,1/Q_s(Y)\\
r_{12}\,\gg\,1/Q_s(Y)
\end{subarray}}
\!\!\!\!\!\!\!\!\!\!\!\!\!d^2 r_2 \,K\Lb r_{01}| r_{12}, r_{02}\Rb\Rb \,\sigma_2\Lb Y,r_{01},b\Rb\nn\\
&+&\,\,
\frac{\bas}{2 \pi} \!\!\!\!\! \int\limits_
 {\begin{subarray}{l}
 r_{02} \,\gg\,1/Q_s(Y)\\
r_{12}\,\gg\,1/Q_s(Y)
\end{subarray}}
\!\!\!\!\!\!\!\!\!\!\!\!\!d^2 r_2 \,K\Lb r_{01}| r_{12}, r_{02}\Rb\,\Big\{  \,\sigma_1\Lb Y, r_{02},b\Rb \, \sigma_1\Lb Y, r_{12},b\Rb\\
 & +  &\,\sigma_2\Lb Y,r_{02},b\Rb \Lb 2\,\Delta_0\Lb Y, r_{12} , b\Rb\,-\,\Delta_{sd}\Lb Y, r_{12} , b\Rb \Rb\,\,+\,\,
 \sigma_2\Lb Y,r_{12},b\Rb \Lb 2\,\Delta_0\Lb Y, r_{02} , b\Rb\,-\,\Delta_{sd}\Lb Y, r_{02} , b\Rb \Rb\nn
  \eea   
In momentum representation in the region with the geometric scaling behaviour \eq{XS21} takes the form:
 \beq \label{XS22}
  \kappa\frac{d \,\sigma_2\Lb \zz\Rb}{d\,\zz}\,\,=\,\, - \zz \,\sigma_2\Lb \zz\Rb\,\,+\,\,2\,\Lb\,\underbrace{  2\,\Delta_0\Lb \zz\Rb\,-\,\Delta_{sd}\Lb \zz\Rb}_{ \Delta_{in}\Lb \zz\Rb} \Rb\sigma_2\Lb \zz\Rb  
  \,+
 \,\,\sigma^2_1\Lb \zz\Rb
  \eeq
 where $\Delta_{in}\Lb \zz\Rb $ is the moment representation of $\Delta_{in}\Lb z\Rb$  with $\sigma_{in}\Lb z \Rb  =  1 - \Delta_{in}\Lb z\Rb$.
 
 The general solution to \eq{XS22} is a sum of  the solution to homogeneous  part of the equation and the particular solution of the non-homogeneous one.  The homogeneous solution has the form:
  \beq \label{XS23} 
 \sigma_2\Lb \zz\Rb\,\,=\,\,\sigma_1\Lb \zz\Rb \exp\Lb- 2\intl^{\infty}_{\zz} \Delta_{in}\Lb \zz'\Rb d \zz'\Rb
 \eeq
 where $\sigma_1\Lb \zz\Rb  $ is given by \eq{XS1MR4}.
 
 We can find the  initial conditions for $\sigma_n$ by  assuming that for $\zz \,<\,0$ only the BFKL 
 Pomeron exchange contribute to $\sigma_1$, while all $\sigma_k =0$  in this kinematic region.  In other words, we assume  that we can neglect  the "fan" BFKL Pomeron diagrams  dealing only with the BFKL Pomeron exchange for $\zz < 0$.  For this initial conditions the solution to \eq{XS22} can be written as:
 \beq \label{XS24}
 \sigma_2\Lb \zz\Rb\,\,\,=\,\,\sigma_1\Lb \zz\Rb \exp\Lb- 2\intl^{\infty}_{\zz} \Delta_{in}\Lb \zz'\Rb d \zz'\Rb
\int\limits^{\zz}_0  \sigma_1\Lb \zz'\Rb\frac{d \zz'}{\kappa},\eeq
 if we neglect the contributions of the order of $\sigma_1^3$.
  
  ~
  ~

 \begin{boldmath}
  \subsection{ $\sigma_3\Lb Y, r,b\Rb$  and $\sigma_k \Lb Y, r,b\Rb $} 
  \end{boldmath}
Using \eq{XSK} we obtain for  $\sigma_3(Y;r_{01},b)$:

    \bea \label{XS30}
   \frac{\partial \sigma_3 (Y;r_{01},b)}{\partial Y} \,\,&=&\,\,
 \frac{\bas}{2 \pi} \!\!\!\!\! \int\limits_
 {\begin{subarray}{l}
 r_{02} \,\gg\,1/Q_s(Y)\\
r_{12}\,\gg\,1/Q_s(Y)
\end{subarray}}
\!\!\!\!\!\!\!\!\!\!\!\!\!d^2 r_2 \,K\Lb r_{01}| r_{12}, r_{02}\Rb \Bigg\{ \sigma_3\Lb Y,r_{02},b\Rb \,+\,\sigma_3\Lb Y,r_{12},b\Rb\,-\,\sigma_3\Lb Y,r_{01},b\Rb\,\,\nn\\
&+&\,\,\sigma_3\Lb Y,r_{02},b\Rb\,\sigma_{sd} \Lb Y,r_{02},b\Rb\,
\,+\,\sigma_3\Lb Y,r_{12},b\Rb\,\sigma_{sd}\Lb Y,r_{02},b\Rb\nn\\
&+& \sigma_2\Lb Y,r_{02},b\Rb\,\sigma_{1}\Lb Y,r_{12},b\Rb 
\,\,+\,\,\, \sigma_2\Lb Y,r_{12},b\Rb\,\sigma_{1}\Lb Y,r_{02},b\Rb \nn\\
&-&\,2\,\sigma_3\Lb Y,r_{02},b\Rb\,N \Lb Y,r_{02},b\Rb\,-\,2\,
\sigma_3\Lb Y,r_{12},b\Rb\,N\Lb Y,r_{02},b\Rb\Bigg\}
\eea
Using the asymptotic expansion for $\sigma_{sd}$ and $N$ we obtain for the  solutions with  geometric scaling behaviour :
 \beq \label{XS31}
  \kappa\frac{d \,\sigma_3\Lb \zz\Rb}{d\,\zz}\,\,=\,\, - \zz \,\sigma_3\Lb \zz\Rb\,\,+\,\,2\,\Delta_{in}\Lb\zz\Rb\sigma_3\Lb \zz\Rb  
  \,+
 \,\,2\sigma_2\Lb \zz\Rb\,\sigma_1\Lb \zz\Rb  \eeq
 
 The solution to \eq{XS31} has the form:
  \beq \label{XS32} 
\sigma_3\Lb \zz\Rb\,\,\,=\,\,\sigma_1\Lb \zz\Rb \exp\Lb- 2\intl^{\infty}_{\zz} \Delta_{in}\Lb \zz'\Rb d \zz'\Rb\Bigg(\int\limits^{\zz}_0 \sigma_1\Lb \zz'\Rb\ \frac{d \zz'}{\kappa}\Bigg)^2\eeq

The general equation for $\sigma_k\Lb Y, r,b\Rb$ reads as follows:
   \bea \label{XSK01}
   \frac{\partial \sigma_k(Y;r_{01},b)}{\partial Y} \,\,&=&\,\,
 \frac{\bas}{2 \pi} \!\!\!\!\! \int\limits_
 {\begin{subarray}{l}
 r_{02} \,\gg\,1/Q_s(Y)\\
r_{12}\,\gg\,1/Q_s(Y)
\end{subarray}}
\!\!\!\!\!\!\!\!\!\!\!\!\!d^2 r_2 \,K\Lb r_{01}| r_{12}, r_{02}\Rb \Bigg\{ \sigma_k\Lb Y,r_{02},b\Rb \,+\,\sigma_k\Lb Y,r_{12},b\Rb\,-\,\sigma_k\Lb Y,r_{01},b\Rb\,\,\nn\\
&+&\,\,\sigma_k\Lb Y,r_{02},b\Rb\,\sigma_{sd} \Lb Y,r_{02},b\Rb\,
\,+\,\sigma_k\Lb Y,r_{12},b\Rb\,\sigma_{sd}\Lb Y,r_{02},b\Rb
+\sum_{j=1}^{k-1}  \sigma_{k - j}\Lb Y,r_{02},b\Rb\,\sigma_{j}\Lb Y,r_{12},b\Rb  \nn\\
&-&\,2\,\sigma_k\Lb Y,r_{02},b\Rb\,N \Lb Y,r_{02},b\Rb\,-\,2\,
\sigma_k\Lb Y,r_{12},b\Rb\,N\Lb Y,r_{02},b\Rb\Bigg\}
\eea

We can check that solution to \eq{XSK01} has the following form:
  \beq \label{XSK02} 
\sigma_k\Lb \zz\Rb\,\,\,=\,\,\sigma_1\Lb \zz\Rb \exp\Lb- 2\intl^{\infty}_{\zz} \Delta_{in}\Lb \zz'\Rb d \zz'\Rb\Bigg(\int\limits^{\zz}_0 \sigma_1\Lb \zz'\Rb\ \frac{d \zz'}{\kappa}\Bigg)^{k-1}\eeq
Note that in term $ \sum_{j=1}^{k-1}  \sigma_{k - j}\Lb Y,r_{02},b\Rb\,\sigma_{j}\Lb Y,r_{12},b\Rb$ in \eq{XSK01} we can replace the factor $\exp\Lb- 2\intl^{\infty}_{\zz} \Delta_{in}\Lb \zz'\Rb d \zz'\Rb \,=\,1$ keeping accuracy of the order of $\sigma^k_1$. 

In \eq{XSK02} we have two smooth functions that have to be determined from the initial conditions: $C(z)$ of \eq{XS1MR4}  and function $D\Lb\zz\Rb$ in the expression for $2 \Delta_{in}\Lb \zz\Rb= D(\zz)\exp\Lb - \frac{\zz^2}{2\,\kappa}\Rb$. We believe that the only condition that we need is
\beq \label{CON1}
\intl^{\infty}_0 C\Lb \zz'\Rb\,\frac{d \zz'}{\kappa}\,=\,\,1
\eeq
In this case 
 the unitary constraints of \eq{UNIT} or/and \eq{INXS1}:
\beq \label{XSK03}
\sigma_{in}\Lb \zz\Rb\,\,=\,\,\sum^{\infty}_{k=1} \,\sigma_k\Lb \zz\Rb\,\,\propto\,\,\zz\eeq
which provides that $\sigma_{in}\Lb z \Rb$ in the coordinate representation reaches the unitarity limit $\sigma_{in}\Lb z \Rb  =  1$. The sum itself has the form:
\beq \label{XSK04}
\sigma_{in}\Lb \zz\Rb\,\,=\,\,\frac{ e^{ - \zz^2/2 \kappa}C\Lb \zz\Rb \exp\Lb - \intl^\infty_{\zz} D\Lb \zz'\Rb e^{ - \zz'^2/2 \kappa}\frac{d \zz'}{\kappa}\Rb}{  \intl^\infty_{\zz} C\Lb \zz'\Rb e^{ - \zz'^2/2 \kappa}\frac{d \zz'}{\kappa}}
\eeq
 
 while
 \beq \label{XSK05} 
 \sigma_k\Lb \zz\Rb\,\,=\,\,e^{ - \zz^2/2 \kappa}C\Lb \zz\Rb \exp\Lb - \intl^\infty_{\zz} D\Lb \zz'\Rb e^{ - \zz'^2/2 \kappa}\frac{d \zz'}{\kappa}\Rb\Lb  1\,\,-\,\, \intl^\infty_{\zz} C\Lb \zz'\Rb e^{ - \zz'^2/2 \kappa}\frac{d \zz'}{\kappa}  \Rb^{k-1}
\eeq 
 Neglecting $  \intl^\infty_{\zz} D\Lb \zz'\Rb e^{ - \zz'^2/2 \kappa}\frac{d \zz'}{\kappa}  \,\sim  e^{ - \zz'2/2 \kappa} 
\,\ll\,1$ and denoting  $\intl^\infty_{\zz} C\Lb \zz'\Rb e^{ - \zz'^2/2 \kappa}\frac{d \zz'}{\kappa}  $ by $1/\bar{n}\Lb \zz\Rb$ we see that
 \beq \label{XSK06} 
 \sigma_k\Lb \zz\Rb\,\,=\,\,e^{ - \zz^2/2 \kappa}C\Lb \zz\Rb \Lb  1\,\,-\,\, \frac{1}{\bar{n}\Lb \zz\Rb} \Rb^{k-1}
 \xrightarrow{\bar{n}\Lb \zz\Rb\,\,\gg\,\,1} e^{ - \zz^2/2 \kappa}C\Lb \zz\Rb \exp\Big( - \frac{k-1}{\bar{n}\Lb \zz\Rb}\Big)
 \eeq

\eq{XSK06} as well as all equations in this paper, are  written at fixed impact parameter $b$. Hence we need to integrate $ \sigma_k \Lb \zz\Rb$ over $b$  to obtain the cross section of produced $k$-gluons.  The $b$-dependence enters our formulae through the saturation momentum (see \eq{z}, \eq{EIGENF1}  and \eq{zm}).  In perturbative QCD the variable $\xi$ in \eq{z}  is  determined  by the form  of  the eigenfunction for  the BFKL equation 
(see  \eq{EIGENF}).
 However, it has been shown \cite{KW1,KW2,KW3,FIIM} that the typical $b$  that contribute in the integrals for cross sections turn out to be of the order of $r  e^{\lambda Y}$  and they lead to the violation of the Froissart theorem\cite{FROI}. Therefore, we need to introduce the non-perturbative corrections at large $b$.  \eq{XSK06} is written in the convenient form  to introduce such corrections to the saturation momentum: $Q^2_s\Lb Y,b\Rb = Q^2_s\Lb Y,b=0\Rb\exp\Lb - \mu\,b\Rb$.
, where $\mu$ is a new non-perturbative parameter.  The multiplicity $\bar{n}\Lb \zz\Rb$ takes the form
\beq \label{NPQS}
\bar{n}\Lb \zz(b)\Rb\,\,=\,\,\Bigg(\frac{C\Lb \infty\Rb}{\zz(b=0) - \mu \,b}\exp\Lb - \frac{\Lb\zz(b=0) - \mu b\Rb^2}{2 \kappa}\Rb\Bigg)^{-1}\,\,\,=\,\,\frac{1}{{\cal A}}\, \exp\Lb  \frac{\Lb\zz(b=0) - \mu b\Rb^2}{2 \kappa}\Rb\eeq

where ${\cal A} $ is a smooth function of $\zz$.

 \begin{figure}
 	\begin{center}
 	\leavevmode
 		\includegraphics[width=8cm]{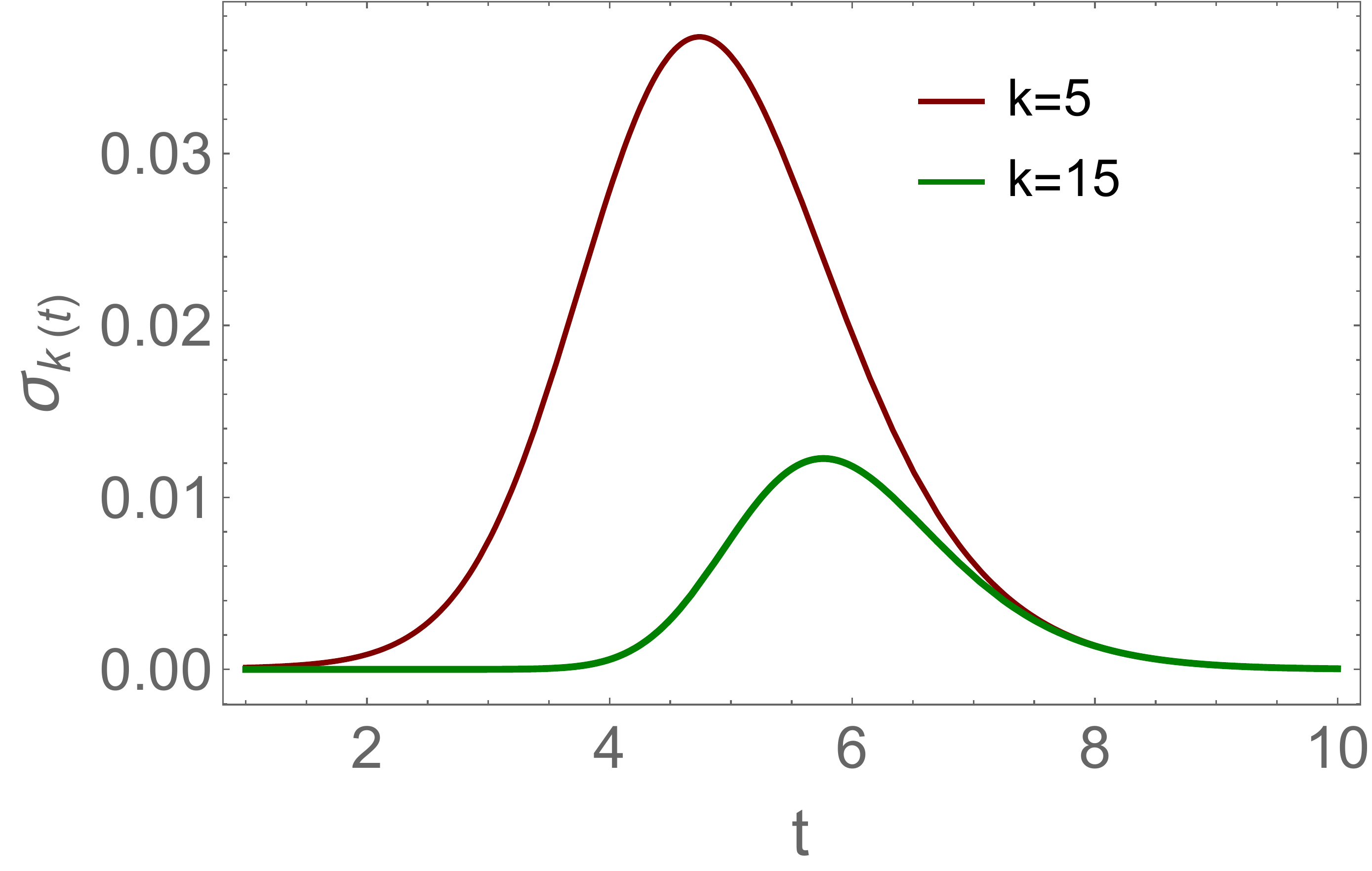}
 	\end{center}
 	\caption{  $\sigma_k\Lb \zz,b\Rb$ in \eq{XSKI1}  versus $t = \zz(b=0) - \mu\,b$ at different values of $k$ . $\kappa=4.88, {\cal A} = 2$.}
 	\label{int}
 \end{figure}

The region of integration over $b$ is $ 0 < b < \zz(b=0)$ since for larger $b$  the scattering amplitude is small and cannot be treated  in our approach. Introducing a new variable $t = \zz(b=0) - \mu\,b$ one can see that for the integral has a maximum at $ t_{SP} =\sqrt{2 \,\kappa\,\ln\Lb {\cal A} (k-1)\Rb} $(see   \fig{int}).. Therefore, for for $t_{SP} \,\leq \zz(b=0)$ the main contribution is determined by the saddle point contribution which gives 
\bea \label{XSKI0}
&&\sigma_k = \\
&&\frac{t_{SP}) (2\,\pi)^{3/2} \kappa}{e\,\mu^2}  \frac{1}{k - 1} \frac{1}{t_{SP}} \,\Big(\zz(b=0) - t_{SP}\Big)=
\frac{  (2\,\pi)^{3/2} \kappa}{e\,\mu^2} \frac{1}{(k-1) }\Bigg(\zz(b=0) -\sqrt{2 \,\kappa\,\ln\Lb {\cal A} (k-1)\Rb}\Bigg)\nn
 \eea
 
 These $\sigma_k$  mildly  depend on $\zz(b=0)$  for all $k-1 \leq \,\bar{n}\Lb \zz(b=0)\Rb$. For larger $k$ the main contributions stem from $b \to 0$.   Estimating the slope at b=0 by the method of steepest decent we obtain  for  $k-1 \,> \,\bar{n}\Lb \zz(b=0)\Rb$ :
\beq \label{XSKI1}
 \sigma_k\Lb \zz(b = 0) ,t=0\Rb\,\,=\,\,\,\intl_{0 < b < \zz(b=0)}\!\!\!\!\!\!\!\! \sigma_k \Lb \zz(b)\Rb d^2 b\,\,=\,\,\frac{2\,\pi}{\mu^2}\,\, \frac{\kappa^2}{\zz(b=0)\,\,\bar{n}\Lb \zz(b=0)\Rb} \frac{\bar{n}^2\Lb \zz(b=0)\Rb}{(k-1)^2}\,\,   \exp\Big( - \frac{k - 1}{\bar{n}\Lb \zz(b=0)\Rb}\Big)\eeq
 
 The notation indicates that the cross section is calculated  for the forward scattering. 
 
 Using \eq{XSKI0} and \eq{XSKI1} we can estimate the value of the total inelastic cross section: 
 \bea \label{TOTIN}
 \sigma_{in}\Lb \zz(b = 0) ,t=0\Rb
 &=& \sum_{k=1}^{\infty}  \sigma_k\Lb \zz(b = 0) \Rb \,\,=\,\,\intl^{\bar{n}\Lb \zz(b=0)\Rb}_1\!\!\!\!\!\! d k \,\sigma_k\Lb \eq{XSKI0}\Rb \,\,+\,\, \intl_{\bar{n}\Lb \zz(b=0)\Rb}^{\infty}\!\!\!\!\!\! d k \,\sigma_k\Lb \eq{XSKI1} \Rb\nn\\
 &=& \frac{(2\,\pi)^{3/2}}{3\,e\,\mu^2}\zz^3(b=0) \,\,+\,\,\frac{2\,\pi}{\mu^2} \frac{\kappa^2}{\zz(b=0)}
  \eea
  
  From \eq{TOTIN} one can see that the main contribution to the total inelastic cross section comes from $k-1 \leq \,\bar{n}\Lb \zz(b=0)\Rb$.

Hence, the probability to emit $k$ gluons at large $\zz$ is equal to

\bea \label{PROB}
p_k\,\,=\,\,\frac{\sigma_k\Lb \zz(b = 0) \Rb}{ \sigma_{in}\Lb \zz(b = 0) ,t=0\Rb}\,\,
=\,\,
 \left\{\begin{array}{l}\,\,\frac{ 2\, \kappa}{\zz^3(b=0) } \frac{\Big(\zz(b=0) -\sqrt{2\,\kappa\,\ln(k-1)} \Big) }{(k-1) }
  \,\,\,\,\,\,\,\,\,\,\mbox{for}\,\,\,k-1 \leq \bar{n}\Lb \zz(b=0)\Rb;\\ \\
\,\,  \frac{2\,e\,\kappa^2}{\zz^4(b=0)} \frac{1}{\bar{n}\Lb \zz(b=0)\Rb} \frac{\bar{n}^2\Lb \zz(b=0)\Rb}{(k-1)^2}\,\,   \exp\Big( - \frac{k - 1}{\bar{n}\Lb \zz(b=0)\Rb}\Big)\,\,\,\,\,\mbox{for}\,\,\, k-1 > \bar{n}\Lb \zz(b=0)\Rb;\\  \end{array}
\right.
 \eea

~

 ~

 \begin{boldmath}
  \section{ Entropy of produced gluons} 
  \end{boldmath}
 
 From \eq{PROB} we can find the value of  von Neumann entropy of produced gluons:
\beq \label{SE}
 S_E\,\,=\,\,-  \sum_{k=1}^{\infty} p_k \ln p_k\,\,\,=\intl^{\bar{n}\Lb \zz(b=0)\Rb}_1\!\!\!\!\!dk \,p_k \ln p_k\,\,+\,\,
 \intl_{\bar{n}\Lb \zz(b=0)\Rb}^{\infty}\!\!\!\!\!dk\, p_k \ln p_k \eeq
 
 From \eq{PROB} one can see that (i) the largest contribution stems from the first sum in  \eq{SE}, since the second term is suppressed as   $1/\zz^4(b=0)$ (see \eq{TOTIN}); and (ii)  in $\ln p_k$ we can restrict 
 ourselves by the term:  $ - \ln(k-1)$.   All other contributions are smaller. Taking the integral we obtain for large values of $\zz(b=0)$:

  \beq \label{SE3}
  S_E\,\,=\,\,0.3\,\frac{\zz^2(b=0)}{2\,\kappa}\,\,  \eeq
  
  In \eq{SE3} we see two strange results. First,  $S_E \propto \frac{\zz^2(b=0)}{2\,\kappa}$.  
   At large $Y$  $\frac{\zz^2(b=0)}{2\,\kappa}$ can be written as $S_E \propto \h \zz \,\bas\,Y$ which is in two times smaller 
  the entropy for the BFKL cascade  estimated in Ref.\cite{KHLE}.
 However, it turns out  that actually extra $\h$ stems from our assumption about geometric scaling behaviour of the cross sections.    In Ref.\cite{KHLE} as well as in Ref.\cite{GOLEE}, it is assumed that the cross sections are functions of $Y$ and $\zz$ but not only $\zz$. Coming back to \eq{XS2} one can see that for $\sigma_1\Lb Y, \zz\Rb$ we obtain the solution: $\sigma_1\Lb Y, \zz\Rb \propto\exp\Lb -\,\bas  Y \zz \Rb$ and  $\bar{n}\Lb \zz(b=0)\Rb  \propto \,\,\exp\Lb \bas  Y \zz \Rb$.  Finally, $S_E \propto \as Y \zz $ in agreement with Ref.\cite{KHLE}. Second, factor $0.3$ in \eq{SE3}  stems from the integration over $b$. Indeed, if we estimate the entropy at fixed $b$ using \eq{XSK06}, we obtain $S_E\,\,=\,\frac{\zz^2(b)}{2\,\kappa}$. Therefore, we can state that non -perturbative corrections that determine the large $b$ behaviour of the saturation momentum, lead to  a decrease of the entropy if we believe in estimate of Refs.\cite{KHLE,GOLEE} for the entropy of the BFKL cascade at the moment of interaction with the target.

   ~
   
   ~
     

  \section{Conclusions} 
 
  
  In this paper we found  the  multiplicity distributions for produced gluons based on the equations derived in Ref.\cite{KLP}. These equations follow from the generating functional approach to Color Glass Condensate effective theory and from AGK cutting rules\cite{AGK}. AGK cutting rules allow us to use the principal property of the BFKL Pomeron: it gives the cross section of produced gluons in  the particular  kinematic region of leading log(1/x) approximation of perturbative QCD.
  
  We solve the equations  in the deep  saturation region where the scattering amplitude has the form of \eq{BKS1}. It turns out that the multiplicity distribution  for $k-1 \,<\, \bar{n} \propto \exp\Lb z^2/2 \kappa\Rb$
   is proportional to  $(1/(k-1))\Big(z- \sqrt{2\,\kappa\,\ln(k-1)}\Big)$ . In this equation
$\bar{n} \propto \exp\Lb z^2/2 \kappa\Rb$   is  the average number of produced  gluons  with $z = \kappa\,\bas Y - \ln Q^2$ where $Q^2$  is the virtuality of the photon and $Y = \ln(1/x)$.  For   $k-1 \,\geq\, \bar{n} \propto \exp\Lb z^2/2 \kappa\Rb$ the multiplicity distribution
 almost reproduces   the KNO scaling behaviour with the average number of gluons $ \bar{n} \propto \exp\Lb z^2/2 \kappa\Rb$. The value of $\kappa$ is given by \eq{GACR}. The KNO function $\Psi\Lb \frac{k}{\bar{n}}\Rb = \exp\Lb -\,k/\bar{n}\Rb$. 
  
  We estimated the value of the entropy of produced gluons which has been a subject of hot recent discussions\cite{KUT,PES,KOLU1,PESE,KHLE,BAKH,BFV,HHXY,KOV1,GOLE1,GOLE2,KOV2,NEWA,LIZA,FPV,TKU,KOV3,KOV4,DVA1}.  We obtain  $S_E= 0.3 \,\frac{z^2}{2 \kappa} $. First factor $ \frac{z^2}{2 \kappa} $ is  
 twice smaller  the entropy of the BFKL cascade estimated in  Ref.\cite{KHLE} (see also Ref.\cite{GOLEE})\footnote{ In Refs.\cite{KOLE,GOLEE} we cut the divergent entropy of the BFKL cascade by the saturation momentum. We believe that the saturation momentum provides the natural scale for interactions with the target.  The result of this paper supports this assumption.}.
 
     As we have discussed in the previous section, we believe that this difference stems from our  assumption that all cross sections should have geometric scaling behaviour. Second, factor $0.3$ in the equation for the entropy stems from the non-perturbative corrections, which provide the correct behaviour of the saturation momentum at large $b$. Therefore, we  claim that the non-perturbative corrections can decrease the entropy if we believe in the estimates of Refs.\cite{KHLE,GOLEE} for the entropy of the BFKL cascade. We would like to stress that we deliberately  calculated the entropy of the produced dipoles,  which can be measured and has been measured in DIS\cite{ENTRDIS}: we evaluated the cross sections  for $n$ produced dipoles, estimate the probability to produce $n$ dipoles ($p_n$)  and determine the entropy $S_E= -  \sum_n p_n  \ln p_n$. It should be noted that the entropy at fixed impact parameter is equal to $ S_E\,\,=\,\,\ln\Lb \bar{n}\Lb z\Rb\Rb = \frac{z(b)^2}{2\,\kappa}$ and coincides with the entropy of the parton cascade estimated in Refs.\cite{KHLE,GOLEE}. Our results can be trusted at large values of $z$ and it is difficult to compare it with the experimental data of Ref.\cite{ENTRDIS}, since it turns out that    for all $x$ in this paper   $z \leq 1$  with the saturation momentum from Ref.\cite{RESH}.

It should be stressed that  we consider DIS in the region of large $z$ where the scattering amplitude has the geometric scaling behaviour. It means that we can discuss DIS with proton at small $x$, but we have to be very careful with DIS on nuclear target.  Indeed, for the nuclear target the geometric scaling behaviour is  valid only in the limited region even at large $z$ \cite{LTHI,KLT,CLM}.  In the vast   part  of the saturation region we do not expect the geometric scaling behaviour for the cross section of produced gluons and have to solve the equation  for $\sigma_k$ with the initial conditions:
\beq 
\sigma_k\Lb \xi, Y=Y_A\Rb= \frac{1}{k!}\Omega^k\Lb Y_A,r,b\Rb \exp|\Lb - \,\Omega\Lb Y_A,r,b\Rb\Rb 
\eeq
where $\Omega\Lb Y_A,r,b\Rb = \sigma_{\mbox{ dipole-proton}} \Lb  Y_A,r\Rb\,T_A\Lb b\Rb$ . $\sigma_{\mbox{ dipole-proton}} \Lb  Y_A,r\Rb  = 2 \int d^2 b N_0\Lb Y_A,r]\Rb$ and $N_0$ is the initial condition for the scattering amplitude of the dipole with the proton. $Y_A = \ln A^{1/3}$ and A is the number of the nucleons in a nucleus. $T_A$   is the optical width of nucleus, which gives the number of nucleons at given value of impact parameter $b$. To find the multiplicity distribution for DIS with nucleus naturally is our  next problem to solve.

   \end{document}